\documentclass[a4paper,11pt]{article}

\usepackage[dvipsnames]{xcolor}
\usepackage[]{color}

\pdfoutput=1 % if your are submitting a pdflatex (i.e. if you have
\usepackage{jcappub} % for details on the use of the package, please
\usepackage[T1]{fontenc} % if needed
\usepackage{aas_macros}

\usepackage{subcaption}
\usepackage{bm}
\usepackage{bbm}
\usepackage{siunitx} % for SI units
\newcommand{\lcdm}{\ensuremath{\Lambda\text{CDM}}}

\newcommand{\one}{\ensuremath{\mathbbm{1}}} % all-ones vector

\newcommand{\dang}{\ensuremath{D_\text{A}}}
\newcommand{\dlum}{\ensuremath{D_\text{L}}}

\newcommand{\Omm}{\ensuremath{\Omega_\text{m}}}
\newcommand{\DD}{\ensuremath{\mathcal{D}}}
\newcommand{\diff}{\ensuremath{\mathrm d}}
\newcommand{\deriv}[2]{\ensuremath{\frac{\diff{#1}}{\diff{#2}}}}
\newcommand{\vect}[1]{\ensuremath{\bm{#1}}} 
\newcommand{\mat}[1]{\ensuremath{\bm{\mathsf{#1}}}}

\usepackage{scalerel}
\usepackage{tikz}
\usetikzlibrary{svg.path}
\definecolor{orcidlogocol}{HTML}{A6CE39}
\tikzset{
  orcidlogo/.pic={
    \fill[orcidlogocol] svg{M256,128c0,70.7-57.3,128-128,128C57.3,256,0,198.7,0,128C0,57.3,57.3,0,128,0C198.7,0,256,57.3,256,128z};
    \fill[white] svg{M86.3,186.2H70.9V79.1h15.4v48.4V186.2z}
                 svg{M108.9,79.1h41.6c39.6,0,57,28.3,57,53.6c0,27.5-21.5,53.6-56.8,53.6h-41.8V79.1z M124.3,172.4h24.5c34.9,0,42.9-26.5,42.9-39.7c0-21.5-13.7-39.7-43.7-39.7h-23.7V172.4z}
                 svg{M88.7,56.8c0,5.5-4.5,10.1-10.1,10.1c-5.6,0-10.1-4.6-10.1-10.1c0-5.6,4.5-10.1,10.1-10.1C84.2,46.7,88.7,51.3,88.7,56.8z};
  }
}
\newcommand\orcid[1]{\href{https://orcid.org/#1}{\mbox{\scalerel*{
\begin{tikzpicture}[yscale=-1,transform shape]
\pic{orcidlogo};
\end{tikzpicture}
}{|}}}}

\title{Non-parametric estimation of the baryon gas fraction and the cosmological bias with clusters}

\author[a]{Hyeon Kim \orcid{0009-0002-3721-3520},}
\author[b]{Raphaël Wicker \orcid{0009-0006-8444-2739},}
\author[a,1]{Benjamin L'Huillier\note{Corresponding author.}\orcid{0000-0003-2934-6243},}
\author[c]{Marian Douspis,}
\author[c]{Laura Salvati,}
\author[d,e]{Arman Shafieloo\orcid{0000-0001-6815-0337}}

\affiliation[a]{Department of Physics and Astronomy, Sejong University, 05006 Seoul, Korea}
\affiliation[b]{Dipartimento di Fisica, Sapienza Università di Roma,
I-00185 Roma, Italy}
\affiliation[c]{Université Paris-Saclay, CNRS, Institut d'Astrophysique Spatiale, 91405, Orsay, France}
\affiliation[d]{Korea Astronomy and Space Science Institute, 
Daejeon 34055, Korea}
\affiliation[e]{University of Science and Technology, 
Daejeon 34113, Korea}

\emailAdd{hyeon970526@sju.ac.kr}
\emailAdd{benjamin@sejong.ac.kr}

\abstract{
X-ray observations of galaxy clusters allow us to estimate the gas fraction, and thus the baryon fraction, and its evolution over time. 
This offers an additional cosmological probe as well as a probe of the gas behaviour in massive halos at the end of structure formation. 
However, cosmological and astrophysical effects are degenerate, and both should be modeled in order to explain observations; otherwise, the chosen baryonic model can potentially bias the cosmological results. 
We propose to quantify this effect by adopting a model-independent framework. 
We utilize Type Ia Supernovae to reconstruct the cosmic expansion history and apply the iterative smoothing method to infer the mass and redshift evolution of the hydrostatic mass bias. 
Our results confirm previous findings and show that the bias should evolve with time to reproduce CMB cosmological constraints. }

\begin{document}

\maketitle
\flushbottom

\section{Introduction}

Galaxy clusters, the most massive gravitationally bound structures in the Universe, serve as powerful cosmological laboratories that provide unique insights into both the evolution of cosmic structure and the fundamental parameters governing our Universe. Their abundance and evolution are exquisitely sensitive to the underlying cosmological model, making them invaluable probes for constraining dark energy, dark matter, and the expansion history of the cosmos \cite{Allen2011, Weinberg2013, 2014A&A...571A..20P,Salvati18,bocquet2024spt}.

Among the various observational approaches to studying galaxy clusters, X-ray observations occupy a central position due to their ability to directly probe the hot intracluster medium (ICM) that dominates the baryonic content of these systems. The X-ray emission from the ICM provides a direct window into the thermodynamic state of the gas, enabling precise measurements of fundamental cluster properties including temperature, density profiles, and crucially, the gas fraction \cite{White1993, Sarazin1988}. The gas fraction, defined as the ratio of gas mass to total gravitating mass within a given radius, represents a key observable that encodes information about both cosmological parameters and the complex astrophysical processes governing baryon physics in massive halos.

The cosmological utility of gas fraction measurements stems from the expectation that  the cosmic baryon fraction should remain approximately constant across different mass scales and redshifts, allowing to get constraints for example on the matter density parameter \Omm\ and the dark energy equation of state \cite{Sadat01,allen2008,Mantz2014}.
However, extracting robust cosmological constraints from gas fraction observations is significantly complicated by different astrophysical processes (baryonic physics) that would impact the gas distribution, such as gas cooling, star formation, supernova feedback, AGN feedback, and gas stripping during cluster assembly \cite{FukugitaPeebles:1998, Crain2007,McCarthy2017,Pillepich2018,angelinelli2023redshift}. 
These astrophysical processes can vary with both cluster mass and redshift, potentially mimicking or masking cosmological signals, and therefore inducing inherent degeneracy between cosmological and astrophysical effects. 

A particularly critical systematic uncertainty arises from the hydrostatic mass bias, which affects the determination of total cluster masses from X-ray observations. This bias, originating from departures from perfect hydrostatic equilibrium due to gas turbulence, bulk motions, and non-thermal pressure support, can significantly impact derived gas fractions and their cosmological interpretation \cite{Rasia2012,Nelson2014}. The magnitude and potential redshift evolution of this bias remain subjects of active investigation, with important implications for cosmological parameter estimation \cite{Salvati19,Pratt2019}.

Traditional approaches to this problem have typically relied on specific parametric models for the various astrophysical processes, introducing potential biases in the derived cosmological constraints depending on the chosen baryonic physics model \cite{Battaglia2013, Le_Brun_2016,wicker2023}. The reliance on particular model assumptions can lead to systematic uncertainties that may be difficult to quantify and could potentially bias cosmological results in subtle but significant ways.

To address these challenges, we propose a novel model-independent framework that avoids the pitfalls associated with specific baryonic physics models while still enabling robust cosmological inference from gas fraction evolution.  
First we study the evolution of the gas fraction as function of mass and redshift, using a smooth reconstruction of cosmic expansion without assuming specific dark energy models, then we move to estimate the evolution of the hydrostatic mass bias  (HMB, hereafter) using smoothing reconstruction instead of a parametric function \cite{wicker2023}. 
In both cases, we keep only the flat Friedmann-Lemaître-Robertson-Walker (FLRW) assumption, and derive evolutions in mass and redshift of the  hydrostatic mass bias independently of the cosmological model.
For that purpose, we apply the iterative smoothing algorithm to the Pantheon+ supernovae data to reconstruct the distance-redshift relation model-independently.
Section~\ref{sec:data} describes the data, models and techniques used in this study, while Section~\ref{sec:res} shows our results on the gas fraction and HMB evolutions. Finally we conclude in Section~\ref{sec:conc}.

\section{Data, Models, and Method\label{sec:data}}
\subsection{Data}

\subsubsection{Observational gas fraction}

Following \cite{wicker2023}, we use the 120 cluster gas fractions of the {\it Planck} ESZ cluster sample \citep{2011A&A...536A...8P} as seen in the follow-up XMM-\emph {Newton} observations analysed in \citep{Lovisari2020}. These clusters span a total mass range from $2.22 \times 10^{14} \mathrm{M_\odot}$ to $1.75 \times 10^{15} \mathrm{M_\odot}$ and redshift range from 0.059 to 0.546. 
The gas fractions were obtained using the gas masses computed from the integrated density in a sphere
\begin{equation}
    \label{eq:mgas_form}
    M_\mathrm{{gas}} (<r) = \int_0^r 4 \pi r'^2 \rho(r') \mathrm{d}r',
\end{equation}
and the hydrostatic masses $M_\mathrm{HE}$ were computed by solving the hydrostatic equilibrium equation,
\begin{equation}
    \label{eq:mhe_form}
    M_\mathrm{HE} (<r) = - \frac{rk_B T(r)}{G \mu m_\text{p}} \left( \frac{\mathrm{d \ln} \rho(r)}{\mathrm{d \ln} r} + \frac{\mathrm{d \ln} T(r)}{\mathrm{d \ln} r} \right),
\end{equation}
where $T$ and $\rho$ are the temperature profile and density profile, $k_B$ is Boltzmann's constant, $G$ is the gravitational
constant, $m_\text{p}$ is the proton mass, and $\mu$ is  molecular weight.

The hydrostatic masses are used here as total masses, although they are known to be biased low due to the hydrostatic equilibrium assumption.
We introduce a factor $B = M_\mathrm{{HE}}/M_\mathrm{{tot, true}}$ to account for this effect.
As a result we define the measured gas fraction as :
\begin{equation}
    \label{eq:fgas_def}
    f_\mathrm{{gas}} = \frac{M_\mathrm{{gas}}}{M_\mathrm{{HE}}} = \frac 1 B \frac{M_\mathrm{{gas}}}{ M_\mathrm{{tot, true}}}.
\end{equation}
Finally, the gas fraction values in the sample span the range $[0.06, 0.20]$. 
The full description of the gas fraction computation is given in \cite{wicker2023} and the complete gas fraction catalog is publicly available in \cite{fgas_catalog_Wicker}. 

\subsubsection{Supernovae data}
\label{sec:sn}

We use the Pantheon+ SNIa compilation \cite{Scolnic2022}, which provides 1701 light-curve entries with distance moduli over $0.01<z<2.26$.
For our analysis we exclude the calibrator subset and all entries with $z<0.01$, leaving $N_\mathrm{LC}=1580$ distance moduli (corresponding to $N_\mathrm{SN}=1466$ unique SNIa) with covariance matrix $\mat{C}_\mathrm{SN}$ including statistical and systematic uncertainties.
The $z<0.01$ cut mitigates sensitivity to peculiar velocities and local bulk flows, which can dominate the observed redshift at very low $z$.
The calibrators are SNIa in hosts with independent geometric distance anchors, primarily intended to set the absolute SNIa magnitude and hence the absolute distance scale in distance-ladder analyses; including them consistently would require an explicit hierarchical calibration model.
Since our goal is to reconstruct the \emph{shape} of the distance--redshift relation (relative distances), with the overall normalization absorbed into nuisance parameters and/or external priors, we remove calibrators and work with the cosmology-only Hubble-diagram sample.
These SNIa are used to reconstruct the distance--redshift relation $\dlum(z)=10^{\mu/5-5}\,\unit{Mpc}$ and the expansion history $h(z)$ directly from the data, with minimal assumptions regarding the cosmological model.

\subsection{Gas Fraction and Bias Models}
\label{sec:model}

In this section, we describe the models assumed by Ref.~\cite{wicker2023} for the gas fraction and the bias. 

The theoretical (hydrostatic) gas fraction is modeled by (Ref.~\cite{allen2008}):
\begin{align}
    f_\text{gas, Th}(M, z) &= K \frac{\Upsilon(M,z)}{B(M,z)} A(z) \left( \frac{\Omega_\text{b}}{\Omega_\text{m}}\right) \left( \frac{\dang^\text{ref}(z)}{\dang(z)}\right)^{3/2} - f_*,
    \label{eq:fgas_model}
    \intertext{where}
     A(z) & = \left( \frac{\theta^\text{ref}_{500}(z)}{\theta_{500}(z)}\right)^{\eta}
     \simeq\left( \frac{H(z)\dang(z)}{[H(z)\dang(z)]^{\text{ref}}}\right)^{\eta}
    \label{eq:A(z)}
\end{align}
is an angular correction and $K$ is an instrumental calibration correction. Regarding the astrophysical contributions, $\Upsilon(M,z)$ is the baryon depletion factor, $ f_*$  is the stellar fraction, and $B(M,z)$ is the hydrostatic mass bias. 
{Finally, the intrinsic scatter of the data $\sigma_f$ is also treated as a free parameter (see eq.~9. in Ref.~\cite{wicker2023}).}
It is worth mentioning here that in addition to the mass bias, the depletion factor $\Upsilon$ may also evolve with mass and redshift, as seen in lower mass systems for instance \cite{Eckert2021}.
However, in the mass range covered by the current cluster sample, such evolutions are not expected from simulation works \cite{planelles2013,Eckert2019_fgas}.

On the cosmological side, ${\Omega_\text{b}}/{\Omega_\text{m}} $ is the universal baryon fraction, $\dang(z) = \dlum(z) / (1+z)^2$ is the angular diameter distance, and $H(z)$ is the Hubble parameter at redshift $z$. 
Therefore, the model of the gas fraction depends on both astrophysical and cosmological assumptions (cosmological model and values of the cosmological parameters).

The bias is modeled assuming both a redshift and a mass evolution:
\begin{align}
     B(M, z) = B_0 \left( \frac{M}{\left< M \right>}\right)^{\alpha} \left( \frac{1+z}{\left< 1+z \right>}\right)^{\beta},
    \label{eq:B(z)}
\end{align}
where $\langle \cdot \rangle$ denotes the mean over the sample.

Ref.~\cite{wicker2023}  estimated  the bias parameters to be $\alpha=-0.057 \pm 0.038  $, $\beta=-0.64 \pm 0.18$, indicating hints for a redshift-evolution of the bias. 
However, as shown in Eq.~\eqref{eq:fgas_model} and~\eqref{eq:A(z)}, the models involve cosmological quantities such as the angular diameter distance $\dang(z)$ and $H(z)$, which are dependent on the underlying cosmological model. 
Therefore, it is important to assess whether or not these results depend on the underlying cosmological model assumption (the \lcdm\ model). 
For that purpose, we will reconstruct distance-redshift relation and expansion history directly from the data, without any cosmological assumptions beyond a flat-FLRW universe.

\subsection{Smoothing Method}
\label{sec:smooth}

To remove cosmological model dependence in Eq.~\eqref{eq:fgas_model}, we have applied the iterative smoothing method \cite{Shafieloo2006,Shafieloo2007,LHuillier2017,Shafieloo2018} to Type Ia supernovae from the Pantheon+ dataset \cite{Scolnic2022}. 

The iterative smoothing method is a data-driven approach that does not rely on a predefined model. 
Starting from various initial conditions $y_0(x)$, as the number of iterations increases, the reconstruction comes closer to the data, converging towards the solution preferred by the data, regardless of its initial conditions. 

\begin{equation}
    \label{eqn:smooth}
    \widehat{y}_{n+1}(x) = {\widehat{y}}_{n}(x) + \frac{\vect{\delta y}^\intercal_{n} \cdot \mat{C}^{-1} \cdot \vect{W}(x)}{\one^\intercal \cdot \mat{C}^{-1} \cdot \vect{W}(x)},
\end{equation}
where
\begin{equation}
    \label{eqn:weight}
    W_i(x)=\exp\left[-\frac{\left({x}-{x_i}\right)^2}{2\Delta^2} \right]
\end{equation}
is the smoothing kernel,  
\begin{equation}
    \label{eqn:residual}
    \left[\vect{\delta y}_{n}\right]_i = y_{i} -  \widehat{y}_{n}(x_i)
\end{equation}
is the data residuals at iteration $n$, 
  $\one^\intercal=(1,\dots,1)$, and $\mat{C}^{-1}$ is the inverse of the data covariance matrix. 
   The weight determines the influence of each data point.   

\subsubsection{Smoothing of the Bias Data}

In the next step, we apply the smoothing algorithm to the bias data. The bias being a function of two variables $(M,z)$, we need to adapt the smoothing algorithm to the two-dimensional case. 
Rewriting Eq.~\eqref{eq:fgas_model} leads to the following form of the bias model:

\begin{align}
    B(M,z)= K \frac{\Upsilon(M,z)}{f_\text{gas}(M, z)+f_*} A(z) \left( \frac{\Omega_\text{b}}{\Omm}\right) \left( \frac{\dang^\text{ref}(z)}{\dang(z)}\right)^{3/2}.  
    \label{eq:bias_model}
\end{align}
In this section, we aim to test the validity of the bias model~\eqref{eq:B(z)}. 

We propagated the uncertainties onto $B$ via a Monte Carlo approach, using priors based on the Planck satellites measurements (see Table~\ref{tab:prior}). 
We then applied a two-dimensional iterative smoothing to the bias data $B(M,z)$.

\begin{align}
    \label{eqn:smooth_2d}
    \widehat{B}_{n+1}(M,z) &  = {\widehat{B}}_{n}(M,z) + \frac{\vect{W}^\intercal(M,z) \cdot \mat{C}^{-1}_\text{B} \cdot \vect{\delta B}_{n}}{ \vect{W}^\intercal (M,z)\cdot \mat{C}^{-1}_\text{B} \cdot\one},\\
    \intertext{where the two-dimensional weight is}
    \label{eqn:weight_bias}
    W_i(M,z) &=\exp\left[-\frac{\ln^2\left(\frac{1+z}{1+z_i}\right)}{2\Delta_z^2} -\frac{\ln^2\left(\frac{M}{M_i}\right)}{2\Delta_M^2}\right],
    \intertext{and the residual vector is}
    \label{eqn:residual_bias}
    \left[\vect{\delta B}_{n}\right]_i & = B_{i} -  \widehat{B}_{n}(M_i, z_i).
\end{align}

The weight must also be treated as a two-dimensional function of mass and redshift. Therefore, the smoothing width must be defined in both mass and redshift directions. As discussed in \cite{Shafieloo2006,Shafieloo2007,LHuillier2017,Shafieloo2018}, the appropriate choice of smoothing width depends on the range and number of data points. In this study, we adopt \((\Delta_M, \Delta_z) = (0.6, 0.2)\), with a number of iterations $N_\text{iter}=2000$. 
Appendix~\ref{app:cv} shows the effect of the choice of the smoothing scales.

\section{Results\label{sec:res}}
\begin{figure}[t]

    \begin{subfigure}[t]{0.32\textwidth}
        \centering
        \includegraphics[scale=0.18]{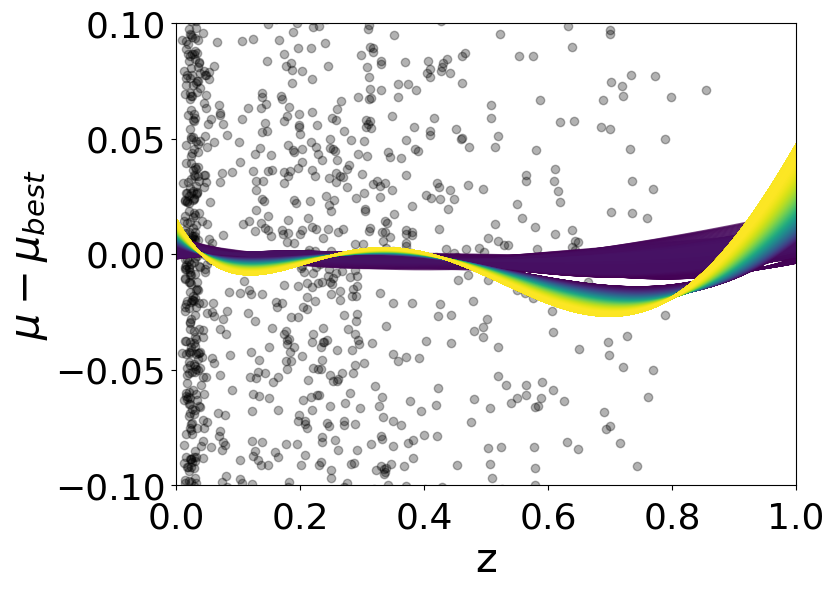}
        \caption{Smooth $\mu$}
        \label{fig:mu}
    \end{subfigure}
    \hfill
    \begin{subfigure}[t]{0.32\textwidth}
        \centering
        \includegraphics[scale=0.18]{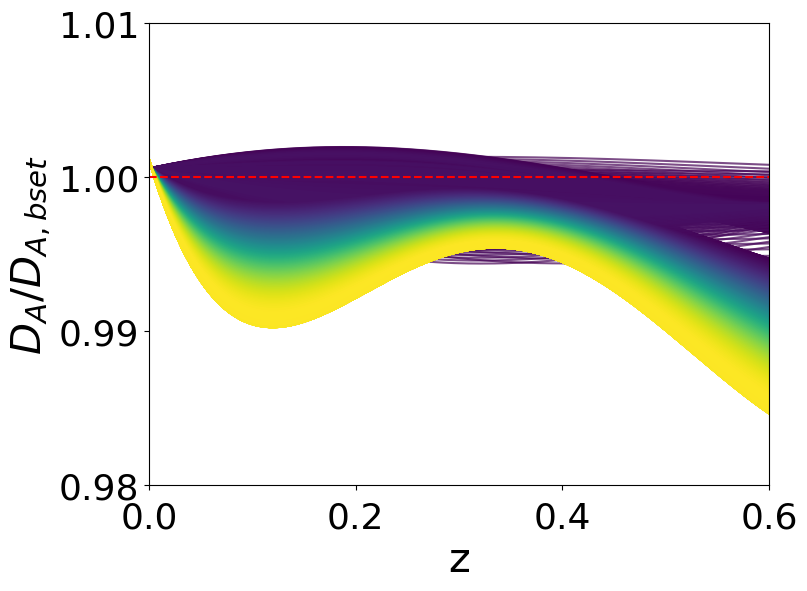}
        \caption{Smooth $\DD$}
        \label{fig:da}
    \end{subfigure}
    \hfill
    \begin{subfigure}[t]{0.32\textwidth}
        \centering
        \includegraphics[scale=0.18]{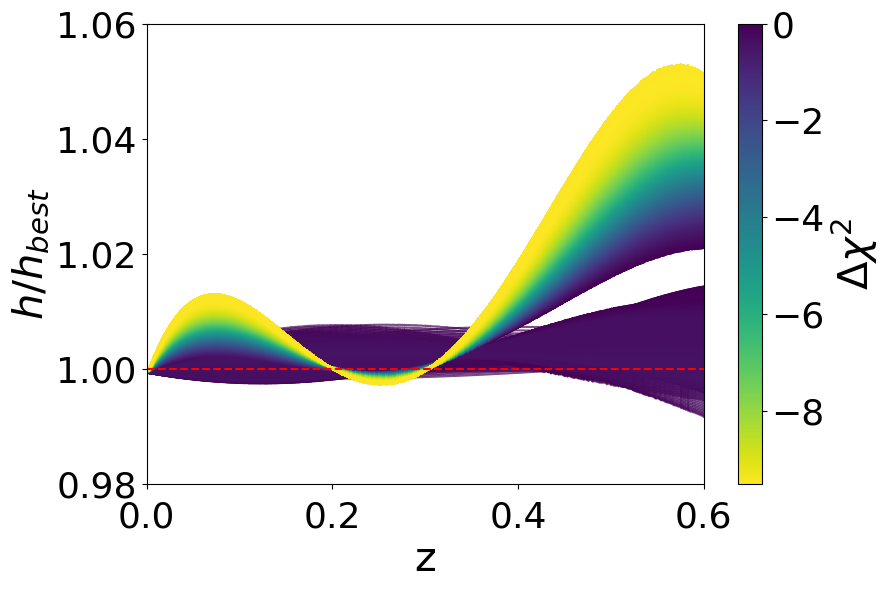}
        \caption{Smooth $h$}
        \label{fig:h}
    \end{subfigure}

    \caption{
   Smooth reconstructions of $\mu(z)$, $\DD(z)$, and $h(z)$ from the Pantheon+ sample. The reconstructions are color-coded by $\Delta\chi^2=\chi^2-\chi^2_{\lcdm}$, where the \lcdm\ best-fit is indicated as a dashed red line.  
    }
    \label{fig:smooth-results}
    
\end{figure}

\subsection{Evolution of the gas fraction}
\label{sec:fgas_smoothdist}

\subsubsection{Sensitivity to expansion history}
 
To reconstruct the distance-redshift without any cosmological assumption aside from a flat Friedmann-Lemaître-Robertson-Walker universe, we apply the iterative smoothing method to the distance modulus $\mu$ from SNIa, using $y= \mu$, $x = \ln(1+z)$, and  $\Delta = 0.3$, as  in previous studies~\cite{Shafieloo2006,Shafieloo2007,LHuillier2017,Shafieloo2018}.
  
We reconstructed the distance modulus starting from  different initial guesses $\hat \mu_0(z)$: Best-fit (to the SNIa data) flat-\lcdm, flat-\lcdm\ with fixed $\Omm\in\{0,0.1,\dots,1\}$, using a number of iteration $N_\text{iter}=8000$. 
We then keep only reconstructions yielding $\chi^2< \chi^2_{\lcdm}$, making it a non-exhaustive set of plausible expansion histories. 

These model-independent reconstructions can then provide the dimensionless comoving distance  
\begin{equation}
    \label{eqn:D(z)}
    \DD(z) = \frac{H_0}{c(1+z)} 10^{\tfrac \mu 5 - 5} \times \SI{1}{Mpc}
\end{equation}
and the expansion history
\begin{equation}
    h(z) = \left[\deriv \DD z\right]^{-1},
    \label{eqn:h(z)}
\end{equation}
assuming a flat universe.

Fig.~\ref{fig:smooth-results} shows the smoothing results of Pantheon+ SNIa and the derived $\mathcal{D}(z)$ and $h(z)$ calculated from Eq.~\eqref{eqn:D(z)} and Eq.~\eqref{eqn:h(z)}.  
In the range $0\leq z \leq z_\text{max} = 0.6$, the smooth distances and expansion histories agree with the best-fit \lcdm\ to $\pm 1\%$ and $\pm 4\%$ respectively, as can be seen from the curves on panels~\ref{fig:da} and \ref{fig:h}.

\begin{figure}[t]
    \centering
    \includegraphics[width=.8\textwidth]{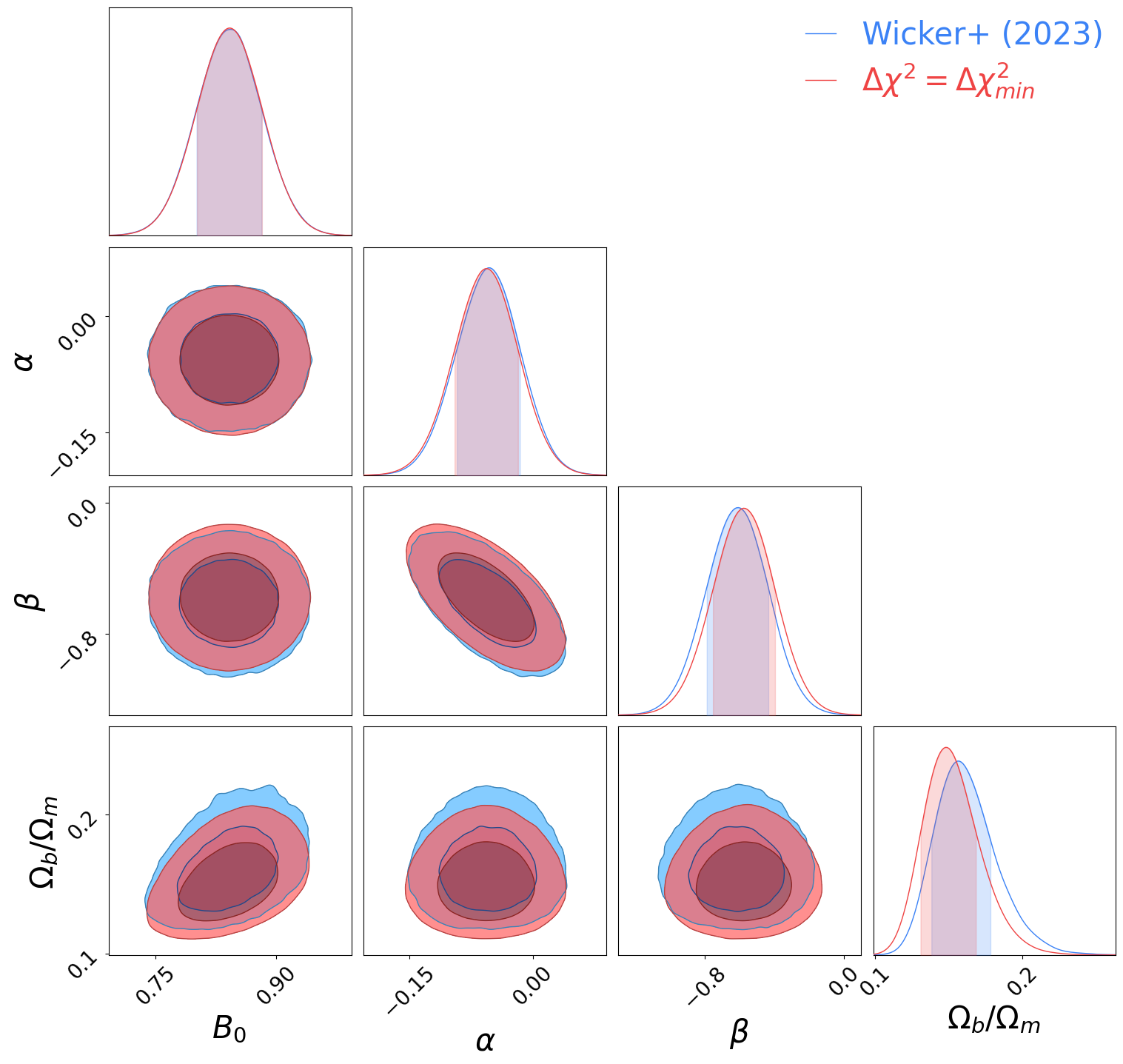}
    \caption{%
    MCMC posteriors for equation \eqref{eq:fgas_model} in the Wicker model (blue) and 
    using smooth reconstructions with $\Delta\chi^2 =\Delta\chi^2_\text{min}=-9.5$ (red). 
    }
    \label{fig:mcmcsmooth}
\end{figure}

In order to test the effect of the expansion history, we then run a Monte-Carlo Markov chain (MCMC) using either the \lcdm\ model for the expansion history, as done in \cite{wicker2023}, or the distances and expansion histories reconstructed via the smoothing method. 
Table~\ref{tab:prior} shows the free parameters of the model and their priors. 
In the smooth expansion case, we do not need $\Omega_\mathrm{m}$, since this term only comes into the expansions and distances in the \lcdm\ case. 
In Fig.~\ref{fig:mcmcsmooth}, the blue contours represent the results from the Wicker model, i.e, eq.~\eqref{eq:fgas_model} with a \lcdm\ expansion. 
The red contours in Fig.~\ref{fig:mcmcsmooth} represent the MCMC results based on the smooth reconstructed expansion history with minimal $\Delta\chi^2$.  
As expected, since the reconstructed expansion histories and distances do not depart significantly from $\lcdm$, both results are consistent. 
In particular, the contours on $\alpha$ and $\beta$ are essentially indistinguishable from the \lcdm\ case. 
In Appendix~\ref{app:iter} we study the effect of the choice the reconstruction on the contours.

\begin{figure}
    \centering
    \includegraphics[width=.8\textwidth]{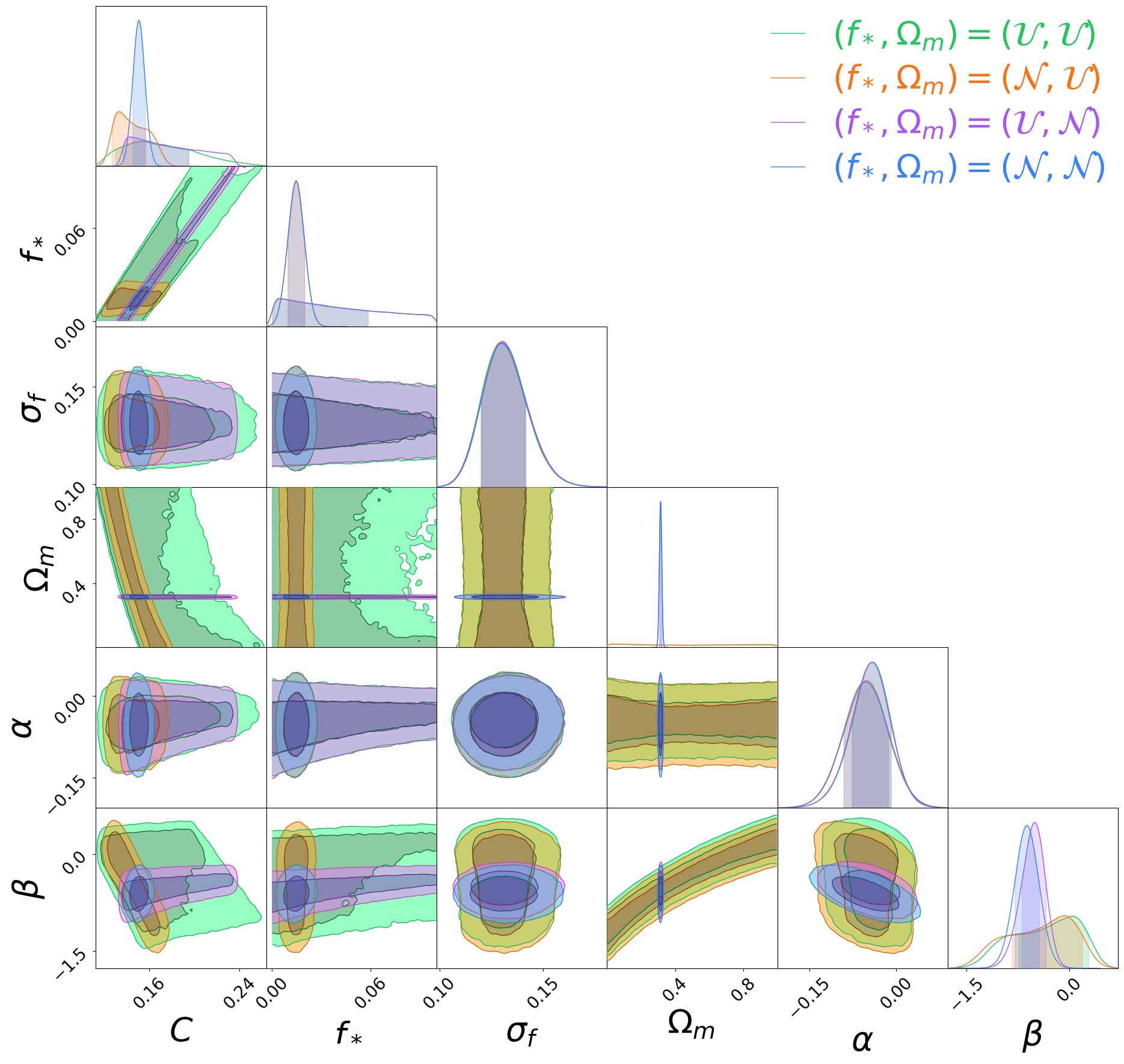}
    \caption{ We present the MCMC results for the 1NP model using normal ($\mathcal N$) and uniform ($\mathcal{U})$ priors, as  specified in Table~\ref{tab:1NP prior}.
    }
    \label{fig:1NPMCMC}
\end{figure}

% 1NP: 
\subsubsection{Effect of priors: 1NP parameterization}
\label{sed:1NP model}

In Eq.~\eqref{eq:fgas_model}, the  parameters $K$, $B_0$, $\Upsilon_0$, $\Omega_\text{b} / \Omm$, and $h$, appear as a simple product, and are therefore degenerate. 
Ref.~\cite{wicker2023} therefore imposed Gaussian priors on these parameters, and the obtained posteriors on these parameters are essentially dictated by the choice of priors. 
To quantify the effects of the choice of Gaussian priors, we combined the parameters
$B_0$, $\Upsilon_0$, $\Omega_\mathrm{b}/\Omm$, and $h$ into a single effective nuisance parameters $C$ defined as:
\begin{align} 
    C& =\frac{K \Upsilon_0} {B_0} 
    \left( \frac{\Omega_\text{b}}{\Omm}\right)
    \left( \frac{h}{0.7}\right)^{3/2}.
\end{align}
We will refer to this model as the One Nuisance Parameter (1NP) model. 

\begin{table}
    \centering
    \renewcommand{\arraystretch}{1.5}
    \caption{Priors used in \cite{wicker2023} and in this work.}
    \label{tab:prior}
    \begin{tabular}{ccccccccccc}
        \hline
		Parameter & Wicker \, VB+OM  & $\textbf{Reference}$  \\ 
		\hline
            $B_0$ & $ \mathcal{U}(0.3,1.7)$ &  - \\
            $f_*$ & $\mathcal{N}(0.015,0.005^2)$ & \cite{eckert2019} \\
            $\Upsilon_0$ & $\mathcal{N}(0.85,0.03^2)$  & \cite{planelles2013} \\
            $K$ & $\mathcal{N}(1,0.1^2)$  & \cite{allen2008} \\
            $\sigma_f$ & $\mathcal{U}(0,1)$  & - \\
            $h$ & $\mathcal{N}(0.674,0.005^2)$   & \cite{planck2020} \\
            $\Omega_b / \Omega_m$ & $\mathcal{N}(0.156,0.003^2)$  & \cite{planck2020} \\
            $\Omega_m$ & $\mathcal{N}(0.315,0.007^2)$  & \cite{planck2020} \\
            $\alpha$ & $\mathcal{U}(-2,2)$  & - \\
            $\beta$ &$\mathcal{U}(-2,2)$  & - \\
		\hline
    \end{tabular}
    \begin{minipage}{0.9\textwidth}
    \footnotesize
    \noindent

    \textbf{Notes:} $\mathcal{U}(a,b)$ denotes a uniform distribution between $a$ and $b$, and $\mathcal{N}(\mu,\sigma^2)$ denotes a normal distribution with mean $\mu$ and standard deviation $\sigma$. 
    \end{minipage}
\end{table}

The priors used in the analysis are the same as those presented in the Wicker 2023 model (Ref.~\cite{wicker2023}), and their details are summarized in Table~\ref{tab:prior}. 
Moreover, the Wicker model employs a very narrow normal prior on the baryon fraction to constrain it. 
Such  tight priors can significantly affect the constraints on $\alpha$ and $\beta$. 
For this reason, in this effective model, we performed analyses using both normal and uniform priors for $f_*$ and $\Omm$, as reported in Table~\ref{tab:1NP prior}.
The results of the MCMC are shown in Fig.~\ref{fig:1NPMCMC}.
The green contours show the results for the uniform priors on $f_*$ and $\Omm$. 
These two parameters are essentially unconstrained, which leads to wide contours on the other parameters, including $C$. 
The orange contours show the posteriors for normal priors on $f_*$ and uniform priors on \Omm. 
The constraints on $\alpha$ and $\beta$ are essentially unchanged, while the constraints on $C$ becomes stronger.
Setting Gaussian priors on \Omm\ while keeping uniform priors on $f_*$, as shown in purple, has the effect of limiting the allowed parameter space in $\beta$, while leaving the posterior on $\alpha$ unchanged. This is because of the strong degeneracy in the $(\Omm, \beta)$ plane. 
Finally, turning both priors to Gaussian, as shown in blue, has the effect of constraining $C$ in addition to $\beta$. 
Therefore, $\alpha$ is relatively independent on the choice of priors on $f_*$ and $\Omm$, while $\beta$ is most sensitive to $\Omm$, and $C$ to the combination of $f_*$ and $\Omm$. 
This analysis confirm the tension found by~\cite{wicker2023} between \Omm\ and no redshift evolution of the bias.

\begin{table}
    \centering
    \renewcommand{\arraystretch}{1.5}
    \caption{Prior used in the 1NP model.  $\mathcal{U}(a,b)$ denotes a uniform distribution between $a$ and $b$, and $\mathcal{N}(\mu,\sigma^2)$ denotes a normal distribution with mean $\mu$ and standard deviation $\sigma$.}
    \label{tab:1NP prior}
    \begin{tabular}{ccccccccccc}
        \hline
		Parameter & Uniform  & Normal  \\ 
		\hline
            $C$ & $ \mathcal{U}(0,10000)$ &  - \\
            $\sigma_f$ & $\mathcal{U}(0,1)$  & - \\
            $\alpha$ & $\mathcal{U}(-2,2)$  & - \\
            $\beta$ &$\mathcal{U}(-2,2)$  & - \\
            $f_*$ & $\mathcal{U}(0,0.1)$ & $\mathcal{N}(0.015,0.005^2)$ \\
            $ \Omm$ & $\mathcal{U}(0,1)$  & $\mathcal{N}(0.315,0.007^2)$ \\
		\hline
    \end{tabular}
\end{table} 

\subsection{ Bias Reconstruction 
}
\label{sec:Bias_sm}
In this section, we analyze the smoothed bias data obtained from Eq.~(\ref{eq:bias_model}) in order to examine the evolutionary behavior of the bias. 
In order to transform a data point $f_b$ at $(M,z)$, one needs to assume a value for the other parameters of the equations as well as an expansion history. 
To do so, we perform a Monte Carlo simulation  on the relevant parameters using the Planck priors from Table~\ref{tab:prior}. 
{The bias data points are obtained by setting the parameter to the central value of the priors, and the errors are calculated by Monte Carlo. 
We then perform the two-dimensional smoothing over the  bias data points obtained. 
The conversion from $f_\text{gas}$ to $B$ necessitate the choice of an expansion history.}
In Section \ref{sec:smbias}, we present the results based on bias data obtained using a \lcdm\ expansion history, while in Section \ref{sec:jointsm}, we analyze the smoothed bias data generated using the reconstructed expansion histories from SNIa.

\subsubsection{Smooth reconstruction of the bias with \lcdm\ expansion}
\label{sec:smbias}

\begin{figure}
    \centering
    \includegraphics[width=1\textwidth]{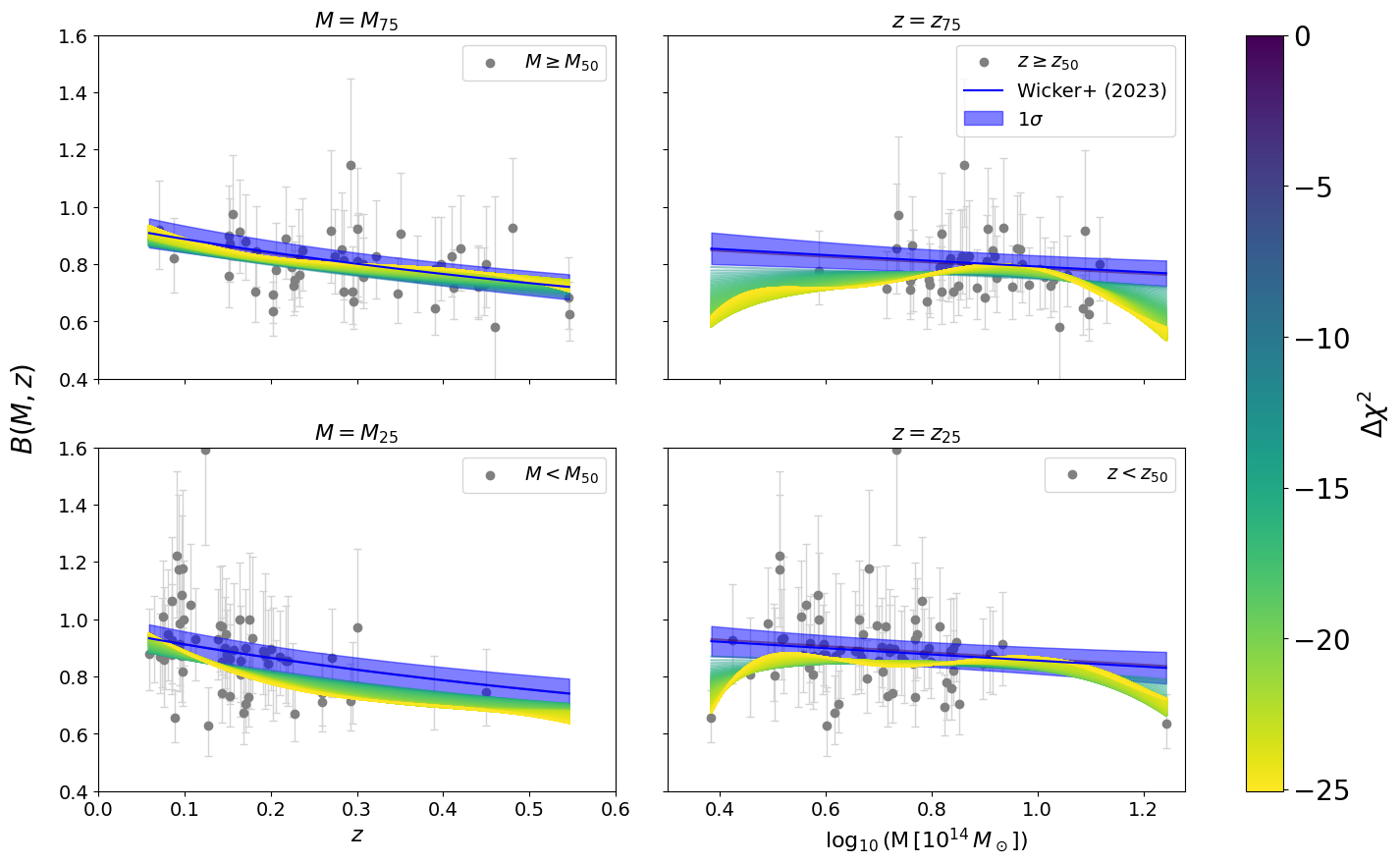}
    \caption{
    Bias data as a function of redshift (left) and mass (right).
    The blue line shows the Wicker et al. best-fit to the data, together with the shaded $1\sigma$ uncertainties.
    The colored solid lines show the reconstructed $\hat{B}(M,z)$ from Eq.~\eqref{eqn:smooth_2d}, color-coded by $\Delta\chi^2$.
    The curves are evaluated at fixed mass slices for the redshift plots (left): $M=M_{75}$ (top-left) and $M=M_{25}$ (bottom-left).
    The curves are evaluated at fixed redshift slices for the mass plots (right): $z=z_{75}$ (top-right) and $z=z_{25}$ (bottom-right).
    The data points correspond to the subsamples $M \geq M_{50}$ (top-left), $M < M_{50}$ (bottom-left), $z \geq z_{50}$ (top-right), and $z < z_{50}$ (bottom-right).
    }
    \label{fig:bias_sm}
\end{figure}

We first reconstruct the bias with eq.~\eqref{eq:bias_model}, using a \lcdm\ expansion history. 
We then apply eq.~\eqref{eqn:smooth_2d} to obtain a smooth reconstruction of $B$ as a function of $z$ and $M$. 

Fig.~\ref{fig:bias_sm} shows the bias as a function of redshift $z$ (left-hand column) and mass $M$ (right-hand column).
To visualize the reconstructed $\hat{B}(M,z)$, we take slices at fixed values of the other variable.
The left panels ($B$ vs $z$) show the reconstruction evaluated at fixed mass $M=M_{75}$ (top) and $M=M_{25}$ (bottom).
The right panels ($B$ vs $M$) show the reconstruction evaluated at fixed redshift $z=z_{75}$ (top) and $z=z_{25}$ (bottom).
Here, $X_{25}, X_{50}, X_{75}$ denote the 25th, 50th, and 75th percentiles of quantity $X$.

The blue line with shaded contours shows the Wicker et al. fit with its $1\sigma$ uncertainties.
The colored solid lines show the smooth reconstruction of the bias (Eq.~\ref{eqn:smooth_2d}), color-coded by $\Delta\chi^2= \chi^2-\chi^2_\text{Wicker}$. 

The data points are similarly divided into subsamples to match the slices: the left panels show data with $M \geq M_{50}$ (top) and $M < M_{50}$ (bottom), while the right panels show data with $z \geq z_{50}$ (top) and $z < z_{50}$ (bottom).
The smooth, non-parametric reconstructions of the bias are consistent with the Wicker model overall, except at the low- and high-mass ranges, where the deviations are driven by a small number of data points.
This overall agreement between the reconstructed bias and the Wicker model serves as a consistency check, validating the parametric form used in previous studies.

\begin{figure}
    \centering
    \includegraphics[width=.9\textwidth]{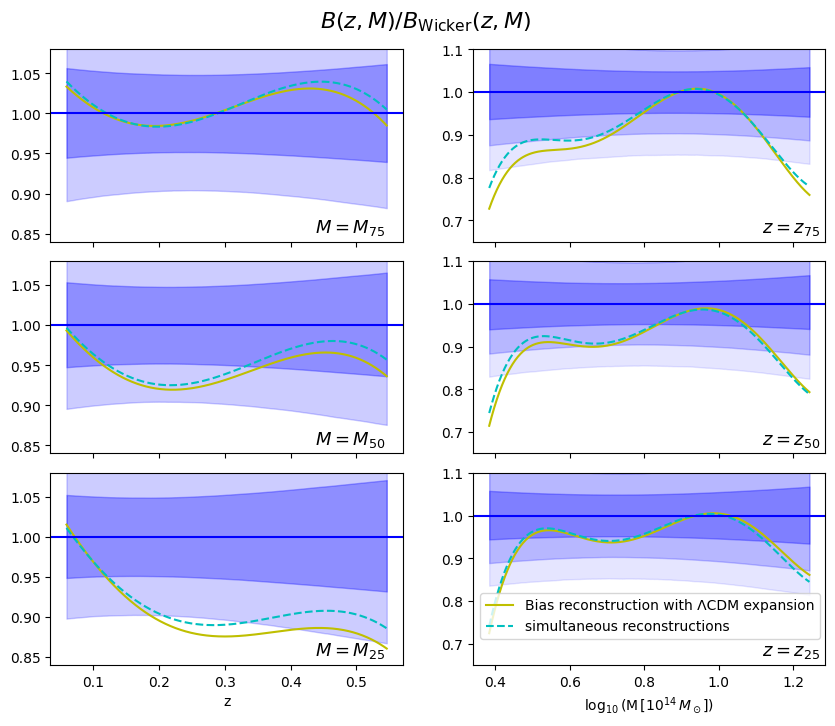}
    \caption{ 
    Reconstructed bias normalized by the Wicker 2023 model. 
    The left column shows the redshift evolution at $M=M_{75}, M_{50},$ and $M_{25}$ respectively in the top, middle, and bottom panels. 
    The right column shows the mass evolution at $z=z_{75}, z_{50}, $ and $z_{25}$ respectively.  
    The olive line (Bias reconstruction) corresponds to the reconstruction based on equation~\eqref{eq:B(z)} using the bias data derived with the Planck prior.  
    The cyan line (Bias + SN reconstruction) represents the reconstruction based on equation~\eqref{eq:B(z)} using bias data inferred from the SN reconstruction.  In both cases, the ratio $\Omega_b / \Omega_m$ is computed using the Planck prior.
    \label{fig:ratio_Bias} 
    }
    
\end{figure}

%%%%%%%%%%%%%%%%%%%%%%%%%%%%%%%%%%%%%%%%%%%%%%%%
\subsubsection{Simultaneous Smoothing of Bias and Expansion}
\label{sec:jointsm}

We now construct the bias data point using eq.~\eqref{eq:bias_model}, with the reconstructed distances and expansions. 
We then apply eq.~\eqref{eqn:smooth_2d} to reconstruct the bias $B(M,z)$. 
We will refer to this approach as ``simultaneous reconstructions'', as opposed to ``bias reconstructions with \lcdm\ expansion'' in \S~\ref{sec:smbias}.

Fig.~\ref{fig:ratio_Bias} shows the reconstruction of $B(M,z)$ with reconstructed expansion histories, normalized by the Wicker model.  
The left-hand column shows the $z$ evolution at $M = M_{75}, M_{50}$, and $M_{25}$ respectively in the top, middle, and bottom rows.
The right-hand columns show the $M$ evolution of the bias at $z = z_{75}$, $z_{50}$, and $z_{25}$ in the top, middle, and bottom rows. 
The blue solid line shows the Wicker (2023) model, and the shaded blue contours show the 1 and 2$\sigma$ contours.
The olive  solid line shows the smooth reconstruction of the bias using the \lcdm\ expansion with the Planck 2018 priors (eq.~\ref{eq:bias_model}), as discussed in S\~ref{sec:smbias}.
We used the iteration with minimal $\chi^2$.

The cyan dashed line shows the reconstruction of the bias using this time the smooth distance reconstructions from SNIa. 
The two lines agree extremely well together, while being consistent with the Wicker model within $3\sigma$, showing the minor effect of the cosmological assumption, and validating the approach.

\section{Discussion and conclusions\label{sec:conc}}

In this study, we perform a consistency test of the assumptions behind the work of \cite{wicker2023} using galaxy cluster baryon fractions as tracers of the evolution of the hydrostatic mass bias. 
In order to remove the cosmological dependence, we perform a model-independent reconstruction of the distance-redshift relation from type Ia supernovae.
First, we test if the constraints on the parameters of the bias evolution found in Ref.~\cite{wicker2023} are robust against the assumption of LCDM. 
We found that given the agreement at low redshift between the LCDM expansion and its reconstruction from  Supernovae the mass and redshift evolutions of the hydrostatic bias are consistent with the original study of Wicker et al. : assuming a Planck prior \citep{2020A&A...641A...6P} on the matter density, the current baryon fractions measured by XMM at low redshifts favor a redshift evolution  of the mass bias at $3.8\sigma$. We show that this result is robust against the choice of criterium to define the ``smoothed'' expansion history.

We then turn to the reconstruction of the evolution of the mass bias without  assuming any parametrized function. For this we adapt and improve our smoothing technic on two aspects. 
We generalize the iterative smoothing in 2 dimensions (mass and redshift) and we marginalize the reconstruction over the cosmological   and astrophysical parameters through MCMC.
Again we compare our result in the \lcdm\  case and in the more general case of flat-FLRW, using the expansion of the Universe derived from supernoave.  We are able to reconstruct the 2D evolution shapes of the bias as function of mass and redshift and to compare it to the parameterization of Wicker et al.  First, our two reconstructions agree with each other as again LCDM being a good fit to the expansion history seen by supernovae. Second, our reconstructions are in agreement within $2\sigma$  with the best fit hydrostatic mass bias evolution found in the original study. The redshift evolution trend found previously is thus confirmed using this new approach.

Although our study shares the same limitations as Wicker et al. the trend we found is consistent with other results including weak lensing studies from Weighing the Giants (WtG, \cite{2014MNRAS.443.1973V}) or the Canadian Cluster Comparison Project (CCCP, \citep{2015MNRAS.449..685H}), showing a mild decreasing trend in mass for the bias for high-z high-mass clusters. Other work, such as the X-ray study
X-COP \citep{2019A&A...621A..40E} or weak lensing study from LoCuSS \citep{2016MNRAS.456L..74S} and CoMaLit
\citep{2017MNRAS.468.3322S}   find trends of a decreasing bias with
redshift on the same mass-redshift range as our sample.

 This study was primarily conducted within a limited redshift range of \( z < 0.6 \), based on a relatively small number of cluster data points. 
 This implies that the analysis was carried out over a narrow cosmological domain, and the smoothing results obtained using the Pantheon+ Ia supernova data are also confined to a region exhibiting only modest variations. 
 Therefore, to more precisely investigate the redshift evolution of the bias, additional analyses based on a broader redshift range and a larger statistical sample are required. 
 New pointed observations of SZ detected clusters (ACT \cite{ACTclusters25}, SPT \cite{SPTcluster25}, SO \cite{SO25}) will soon provide numerous and accurate baryon fractions up to redshift $\sim 1$. In parallel, Supernovae are already providing expansion history at higher redshifts but complementary measurements will come from gravitational waves observations very soon.

\acknowledgments
This work was partially supported by the ``PHC STAR'' programme (project number: 50123WB, RS-2023-00259422), funded by the French Ministry for Europe and Foreign Affairs, the French Ministry for Higher Education and Research, and the National Research Foundation of Korea.
B.~L. acknowledges the support of the National Research Foundation of Korea (NRF-2022R1F1A1076338) and the support of the Korea Institute for Advanced Study (KIAS) grant funded by the government of Korea.

\bibliographystyle{JHEP}
\bibliography{biblio}

\appendix

\section{Effect of smoothing iterations}
\label{app:smoothing_effect}

\subsection{Convergence of the smoothing procedure}
\label{app:cv}
\begin{figure}[t]
    \begin{subfigure}[t]{0.48\textwidth}
        \centering
        \includegraphics[scale=0.5]{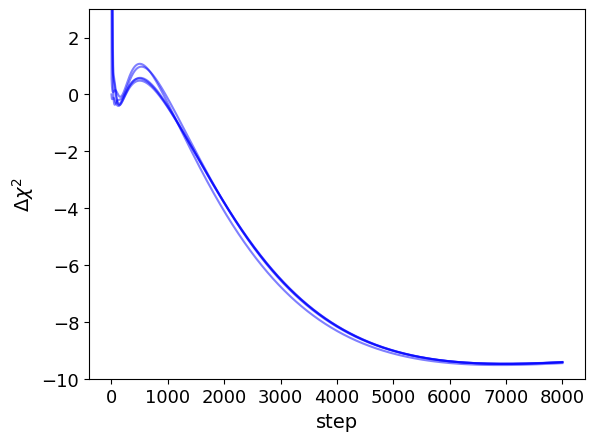}
        \caption{Pantheon+ Smoothing}
        \label{fig:chi2_SN}
    \end{subfigure}
    \hfill
    \begin{subfigure}[t]{0.48\textwidth}
        \centering
        \includegraphics[scale=0.5]{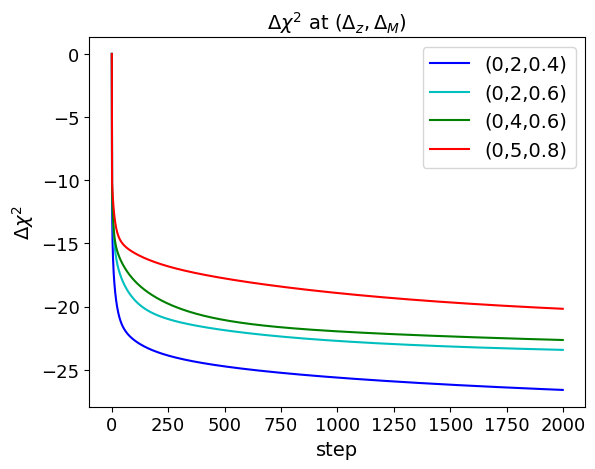}
        \caption{Bias Smoothing}
        \label{fig:chi2_Bias}
    \end{subfigure}
    \caption{The figure shows the evolution of $\Delta \chi^2$ as the smoothing step. The left panel presents the SN smoothing results obtained from multiple initial conditions, while the right panel shows the bias smoothing results for various $(\Delta_z, \Delta_M)$ combinations.
    }
    \label{fig:chi2}
\end{figure}

Fig.~\ref{fig:chi2_SN} shows the evolution of $\Delta\chi^2$  with iteration of the SNIa smoothing, for the different initial guesses. 
All initial guesses yield similar trajectories, with a bump around iteration 1000, then a decreasing behaviour. 
We stopped the iteration at iteration 8000, where $\Delta\chi^2$ seems to slow down.
Fig.~\ref{fig:chi2_Bias} shows the evolution $\Delta\chi^2$ of the bias smoothing (Section~\ref{sec:Bias_sm} for four choices of $(\Delta_z, \Delta_M)$. 
All settings give similar shapes. smalller values of the smoothing parameters give a lower $\Delta\chi^2$ as expected.

\subsection{Effect of the smoothing iteration on the MCMC parameter estimation}

\label{app:iter}

\begin{figure}
    \centering
    \includegraphics[width=.8\textwidth]{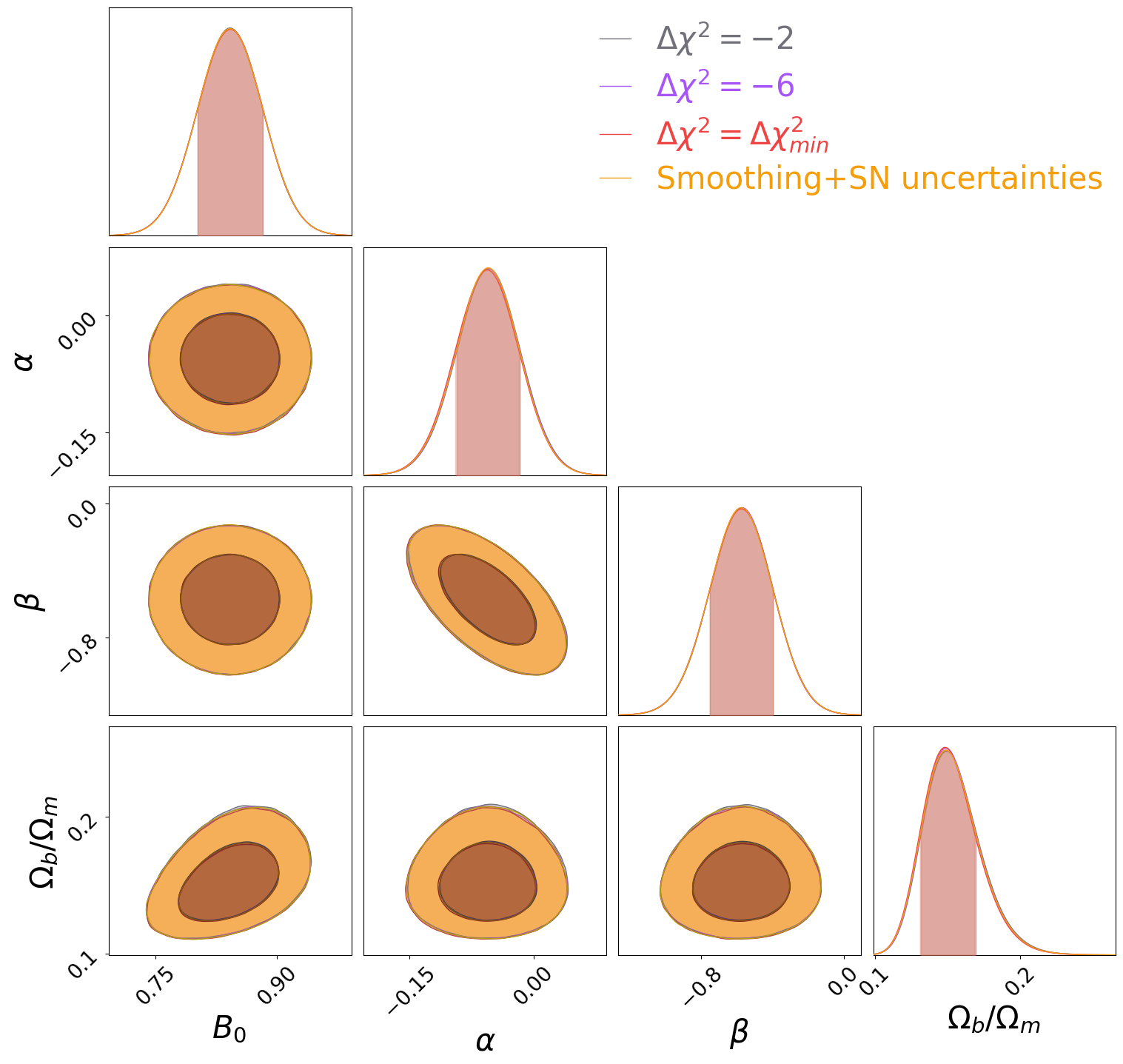}
    \caption{
    Same plot as Fig.~\ref{fig:mcmcsmooth} with other smooth expansion histories corresponding to different  $\chi^2$.
    }
    \label{fig:MCMC_chi2}
\end{figure}

Fig.~\ref{fig:MCMC_chi2} shows the effect of the smoothing iteration on the MCMC. 
We used four different reconstructed $\{h(z), \DD(z)\}$ corresponding respectively to $\Delta\chi^2 = -2, -6, \Delta\chi^2_\text{min}$. 
The posterior distribution of $B,\alpha,\beta,\Omega_b / \Omega_m$ are undistinguishable in all  cases, showing the robustness of the results. 
This is expected, given the tight constraints from the SNIa over the range of redshift of interest.

Finally, in the ochre contour, instead of performing the MCMC at fixed reconstruction, we randomly sampled among the various reconstructions, in order to reflect the uncertainties in the smooth reconstructions. 
The results are again undistinguishable, reflecting the tight constraints of the distances and expansion histories from the SNIa in this redshift range. 

\end{document}